\pdfoutput=1 

\documentclass[10pt,journal,compsoc]{IEEEtran}
\usepackage{listings}
\usepackage{color}
\usepackage{amsmath}

\definecolor{dkgreen}{rgb}{0,0.6,0}
\definecolor{gray}{rgb}{0.5,0.5,0.5}
\definecolor{mauve}{rgb}{0.58,0,0.82}
\ifCLASSOPTIONcompsoc
  \usepackage[nocompress]{cite}
\else
  \usepackage{cite}
\fi
\usepackage{graphicx}
\usepackage{multirow}

\ifCLASSINFOpdf
\else
\fi
\begin{document}
%
\title{Customizable Precision of Floating-Point Arithmetic with Bitslice Vector Types}
%
%
%
%

\author{Shixiong~Xu,
        and~David~Gregg
\IEEEcompsocitemizethanks{\IEEEcompsocthanksitem S. Xu is with the School
of Computer Science and Statistics, Trinity College Dublin, the University of Dublin, Dublin, Ireland.\protect\\
E-mail: xush@tcd.ie
\IEEEcompsocthanksitem D. Gregg is with the School
of Computer Science and Statistics, Trinity College Dublin, the University of Dublin, Dublin, Ireland.\protect\\
E-mail: dgregg@cs.tcd.ie}
}

\IEEEtitleabstractindextext{%
\begin{abstract}
  Customizing the precision of data can provide attractive trade-offs
  between accuracy and hardware resources. We propose a novel form of
  vector computing aimed at arrays of custom-precision floating point
  data. We represent these vectors in bitslice format. Bitwise
  instructions are used to implement arithmetic circuits in software
  that operate on customized bit-precision. Experiments show that this
  approach can be efficient for vectors of low-precision custom
  floating point types, while providing arbitrary bit precision.
\end{abstract}

}

\maketitle

\IEEEdisplaynontitleabstractindextext

%
\IEEEpeerreviewmaketitle

\IEEEraisesectionheading{
\section{Introduction}\label{sec:introduction}}
\IEEEPARstart{O}{ne} of the most important developments over the last decade has been
the move from desktop computing to battery-powered computing in
hand-held, wearable and mobile devices. This move from the desktop to
the wider world is also reflected in the growth of applications that
operate on real world data such as images, video, sound and motion.
These applications are highly computationally intensive, and pose huge
challenges both for mobile devices and for cloud-based services that
receive and process large amounts of such data.

\textit{Approximate computing} can be an effective technique both for
accelerating these types of applications and for reducing the required
energy. Approximate computing is based on the observation that the
inputs and outputs of these algorithms are approximations.
Introducing additional imprecision in the computation may have little
or no effect on the final result.


One of the popular ways of approximate computing is to reduce the
data precision such as from single precision binary-32 floating point
to half precision binary-16. When designing custom hardware, data precision can be customized
precisely to the needs of an application. For example, it has been
found that some applications can make good use of as little as 8-bit
floating point. When designing a custom FPGA or ASIC solution, the
hardware can implement the exact level of required precision.
Reducing precision can reduce the size of the hardware, but crucially
it can also allow less data to be transfered between the processor
and memory. 


On general-purpose processors it is much more difficult to customize
the precision of data to the application. Most general-purpose
processors provide only two floating point sizes --- single and double
precision --- and a limited range of integer data sizes, typically 8,
16, 32, and 64-bit. For example, if 9 bits of integer precision are
required for an application, the programmer will normally use a 16-bit
type.  Similarly, if one needs 13-bit floating point (FP), one might
use a half precision \textit{binary-16} FP type if it is available, or
single precision \textit{binary-32} if not.

When the data consists of a large array of values, the cost of using
more precision than necessary can become large. The obvious problem is
that the larger data size requires more space in memory. But the
larger data size also requires more memory bandwidth when transferring
between processor and memory, and more energy to drive external pins,
wires and buses when transferring unnecessarily large data.

In this paper we propose an entirely novel approach to supporting
arrays of irregular precision floating point data types.
We adopt the idea of \textit{software bitslice} data representations
that are used in the implementation of cryptography algorithms and
some image representation formats. We use these software bitslice
formats to represent arrays of data, and perform vector-style SIMD
operations constructed from simple bitwise logical operations.

We make several contributions:
\begin{itemize}
	\item We propose using software bitslice data representations to
	create a new approach to SIMD vector computation for
	customizable precision floating point data types.
	\item We present our customizable precision bitslice floating point operations as intrinsic functions similar to SIMD intrinsics and implement the operations as a reconfigurable library. 
\item Experimental results show that our bitwise vector approach is
  efficient for large vectors of customized floating point types with
  low precision.

\end{itemize}


%
%
%
%

\section{Software Bitslice Representations}
In the standard representation of simple types, such as integer and
floating point values, a single value fits inside an 8, 16, 32 or
64-bit word. In a bitslice representation, the different bits of a
single number are spread across multiple machine words.

\begin{figure}[!t]
\centering
\includegraphics[width=0.48\textwidth]{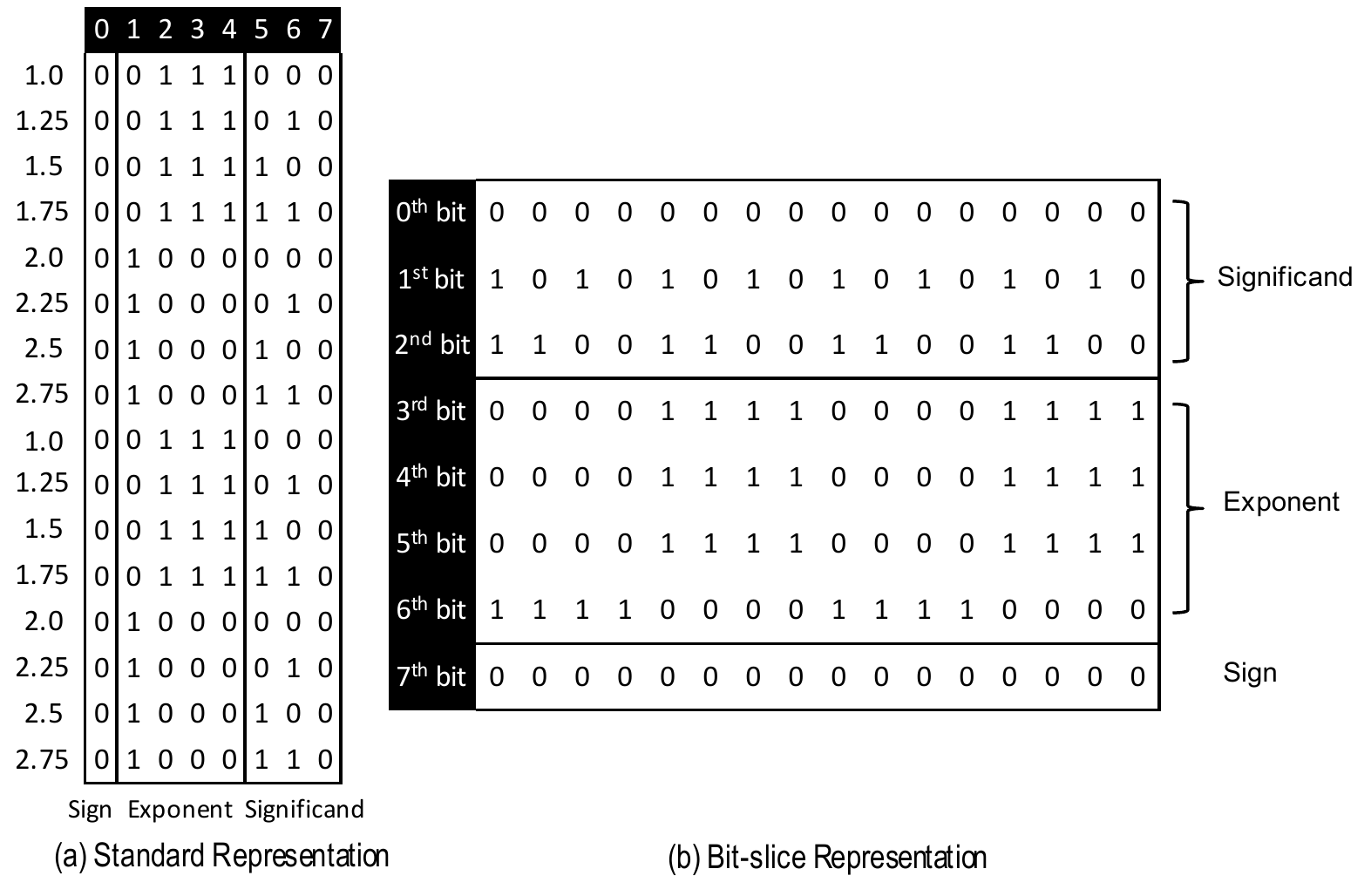}
\caption{Standard and bitslice representation of an array of sixteen 8-bit floating point numbers.}
\label{fig_presentation}
\end{figure}

Figure \ref{fig_presentation} shows an example of standard and
bitslice representations of arrays. Both the standard and bitslice
representations show an array of sixteen 8-bit floating point numbers.
Each number has 1 sign bit, 4 exponent bits and 3 mantissa bits. However,
the physical representation of the data in memory is quite
different. Instead of using sixteen 8-bit words, the bitslice
representation uses eight 16-bit machine words.  The first bit of each
of these eight 16-bit words corresponds to one of the eight bits of
the first array element. Similarly, the other 8-bit values are
represented by a bit from each of the eight 16-bit words.

Bitslice representations are sometimes used for highly efficient
implementations of symmetric cryptography algorithms such as the Data
Encryption Standard (DES) or Advanced Encryption Standard
(AES)~\cite{DES-bitslice}. These cryptography algorithms perform large
numbers of bit-level operations, which can be very fast on bitslice
representations. But to our knowledge they have not been applied to
more general purpose computation.

\section{Bitslice vector computing}
\label{sec:bfp-computing}
In this paper we propose using bitslice representations of arrays as
the basis of a new approach to vector SIMD computing.  We show how to
construct vector SIMD operations in software that operate on these
types. This allows us to construct SIMD vector types of a fixed number
of elements but with arbitrary length per element. For example, we can
construct a vector of thirty-two 8-bit elements, but equally we can
construct a vector of thirty-two 17-bit elements.

\begin{figure}[h]
	\centering
	\begin{lstlisting}
    #define ADD_BITS 13
    uint32_t ary1[ADD_BITS];
    uint32_t ary2[ADD_BITS];
    uint32_t result[ADD_BITS];
    
    uint32_t carry = ALLZEROS;
    for (int i = 0; i < ADD_BITS; i++) {
      t1 = ary1[i];
      t2 = ary2[i];
      xxor = t1 ^ t2;
      aand = t1 & t2;
      // add
      result[i] = xxor ^ carry;
      // update carry
      carry = (carry & xxor) | aand;
    }
	\end{lstlisting}
	\caption{Bitslice adder for two arrays of unsigned integers. Each integer has ADD\_BITS bits. The size of uint32\_t decides the number of array elements are being processed.}
	\label{fig:bitslice-adder}
\end{figure}

To operate on the elements of bitslice vector types, we propose
building arithmetic and other operators from native integer bitwise
instructions. Figure \ref{fig:bitslice-adder} shows a simple integer
adder routine for bitslice vectors with thirty-two elements, each of
13 bits. Note that the addition is performed by a sequence of bitwise
operations that are the software equivalent of a hardware adder. Thus
the sum of two bits $t1$ and $t2$ is $t1 \text{ XOR } t2$ and the
carry from the addition is $t1 \text{ AND } t2$. By applying a
sequence of these bitwise operations, an entire $k$-bit addition can
be performed.

Note that in a hardware adder, each logic gate operates on one binary
value. However, the bitwise logical operators in the adder in
Fig. \ref{fig:bitslice-adder} operate on an entire 32-bit register of values at once.
Thus, the addition is performed sequentially by a sequence of bitwise
operations. But each bitwise instruction operates on 32 separate
13-bit values. So our adder operates in vector SIMD style, requiring a
number of steps that is proportional to the number of bits in each
value, but operating on a vector of different values that is equal to
the word-size of the underlying type supported by the machine.

The big advantage of our proposal for bitslice vector types is that
they allow vectors of values with an arbitrary number of bits. One can
easily support vectors of numbers with 5, 9, or 13 bits. Operating on
bitslice vector types is laborious from a sequential point of view,
but exploits large amounts of bit-level parallelism within the
conventional machine word. The major downside of operating on bitslice
vector types is that each operation requires large numbers of bitwise
operations. As the number of bits in each value grows, the execution
time of the arithmetic operators increases rapidly. However, as we
show in the following sections, it can work well for arrays of small,
irregularly-sized types.

It has been demonstrated that not all programs need the precision
provided by the generic FP hardware and different sections of a
program can benefit from different bitwidths for the sake of overall
accuracy and power consumption \cite{Tong:2000:RPO:340730.340741}.
The balance between accuracy and performance makes our
solution perfectly suited to the needs of approximate computing.




\section{Bit-slice Floating Point Vector Operations}

Bitslice floating point (BFP) vector operations perform arithmetic computations on the bitslice vector types. Bitslice vector types are essentially an array of unsigned integers (e.g., uint8\_t, uint16\_t, and \_\_m128i in Intel SSE instructions), each of the integers represents one bit of the data in the standard representation. Figure \ref{fig:bitslice-types} shows the BFP vector type for FP32. The width of bitslice vector types is decided by the size of the underlying integer types. For example, for uint16\_t, the width is 16. As discussed in Sec. \ref{sec:bfp-computing}, the arithmetic operations on bitslice vector types are carried out in terms of a single bit rather than the whole value. Therefore, we need use integer bitwise operations to achieve the logic of hardware for each arithmetic operation. Our implementation follows the classic implementation of floating arithmetic operations in hardware ~\cite{opac-b1130488} but with the aim of minimizing the number of gates rather than the overall latency. 

\begin{figure}[h]
	\centering
	\begin{lstlisting}
	typedef uint16_t BFP_ELEM_TYPE;
	#define SIGN_BIT 1
	#define EXPO_BIT 8
	#define SIG_BIT 23
	typedef struct{
	BFP_ELEM_TYPE data[SIGN_BIT + EXPO_BIT + SIG_BIT];
	} BFP_FP_VEC_TYPE;
	
	\end{lstlisting}
	\caption{Bitslice floating point vector types for FP32.}
	\label{fig:bitslice-types}
\end{figure}

In addition to the helper operations for transforming data in the standard format to our BFP vector types, we give three basic arithmetic operations -- addition/subtraction, multiplication, division. For each operation, two rounding modes are available -- \emph{round towards zero} and \emph{round to nearest} (tie to even). The computation steps and associated complexities for each BFP operations are listed in Table \ref{tab:computation-steps}. For the division, we adopt the \emph{restoring division algorithm}, which is the simplest digit-recurrence algorithm~~\cite{MullerEtAl2010}.

\begin{table}[!bhpt]
	\caption{Computation steps and complexities for BFP arithmetic operations. The complexity of rounding is for round towards zero.}
	{\begin{tabular}{|c|c|c|c|c|c|}
			\hline   &  step1 & step2 & step3 & step4  & step5 \\ 
			\hline  \multirow{ 2}{*}{Add/Sub}  &  sign & align. shift & add & normal. & round\\ 
			\cline{2-6}    &  $O(1)$ & $O(nlogn)$ & $O(n)$ & $O(nlogn)$ & $O(n)$\\ 
			\hline  \multirow{ 2}{*}{Mul}      &  sign & add expo  & mul & normal. &  round\\ 
			\cline{2-6}        &  $O(1)$ & $O(n)$  & $O(n^2)$ & $O(nlogn)$ &  $O(n)$ \\ 
			\hline  \multirow{ 2}{*}{Div}      &  sign & sub expo &  normal.  & div &  round\\ 
			\cline{2-6}        &  $O(1)$ & $O(n)$ &   $O(n)$   & $O(n^2)$  &  $O(n)$\\ 
			\hline 
		\end{tabular}}
		
		\label{tab:computation-steps}
	\end{table}

Bit shifting is required in the alignment shift of the add operation and normalization of all the operations. In BFP vector types, bits of a vector item are spread over different integers. Shifting bits one by one is thus prohibitive due to the memory access in proportion to the number of bits to be shifted. We adopt a log shifter that significantly reduces the total number of bits to be shifted and in turn eliminates some unnecessary memory access. Log shifting has been demonstrated as an effective way of saving power in hardware~\cite{Pillai:1997:EDM:263272.263341}\cite{Acken:1996:PCB:252493.252605}. 

Our BFP vector types are presented in the form of arrays and thus it is of great importance to exploit the data locality within each computation step (e.g. loops) and across steps. For example, software pipelining with loop unrolling is applied to the multiplication of significand in order to keep data in registers, and reuse them as much as possible. For similar operations, such as two successive addition operations on exponent, we can merge them together to avoid the store of intermediate results.

\section{Experimental evaluation}

\subsection{Implementation}
Our proposed customizable precision floating-point arithmetic is implemented as a reconfigurable bitslice FP library. When the programmer knows the bitwidth requirement for their applications in advance, they can simply put the number of bits in exponent, mantissa, and rounding mode into a configuration file and feed it to our library generator. The generator produces a header file containing the BFP data structures and related C intrinsic functions for the basic BFP operations , and a library file (.so) that implements the custom FP operations. Programmers can either manually modify their code with BFP vector types and operations or annotate the source code and let compilers vectorize the their code and automatically generate BFP operations. The compiler support is beyond the scope of this paper.


\subsection{Experimentation Evaluation}

We evaluated the performance of our BFP vector operations on a Linux platform with an Intel(R) Core(TM) i7-4770 CPU,  which supports AVX2 SIMD instructions. We used BFP vector types to represent three floating point formats -- FP8 (1 sign bit, 4 exponent bits, 3 significand bits), FP 16 (1 sign bit, 5 exponent bits, 10 significand bits), and FP32 (1 sign bit, 8 exponent bits, 23 significand bits). For each format, we measured the performance with different integer types supported by the CPU, from 32-bit integer (uint32\_t) to 256-bit integer(\_\_m256i). As our implementation supports two widely used rounding modes -- round towards zero and round to nearest, performance is given for comparison as well.
  \begin{figure}[!bhpt]
  	\centering
  	\includegraphics[width=0.45\textwidth]{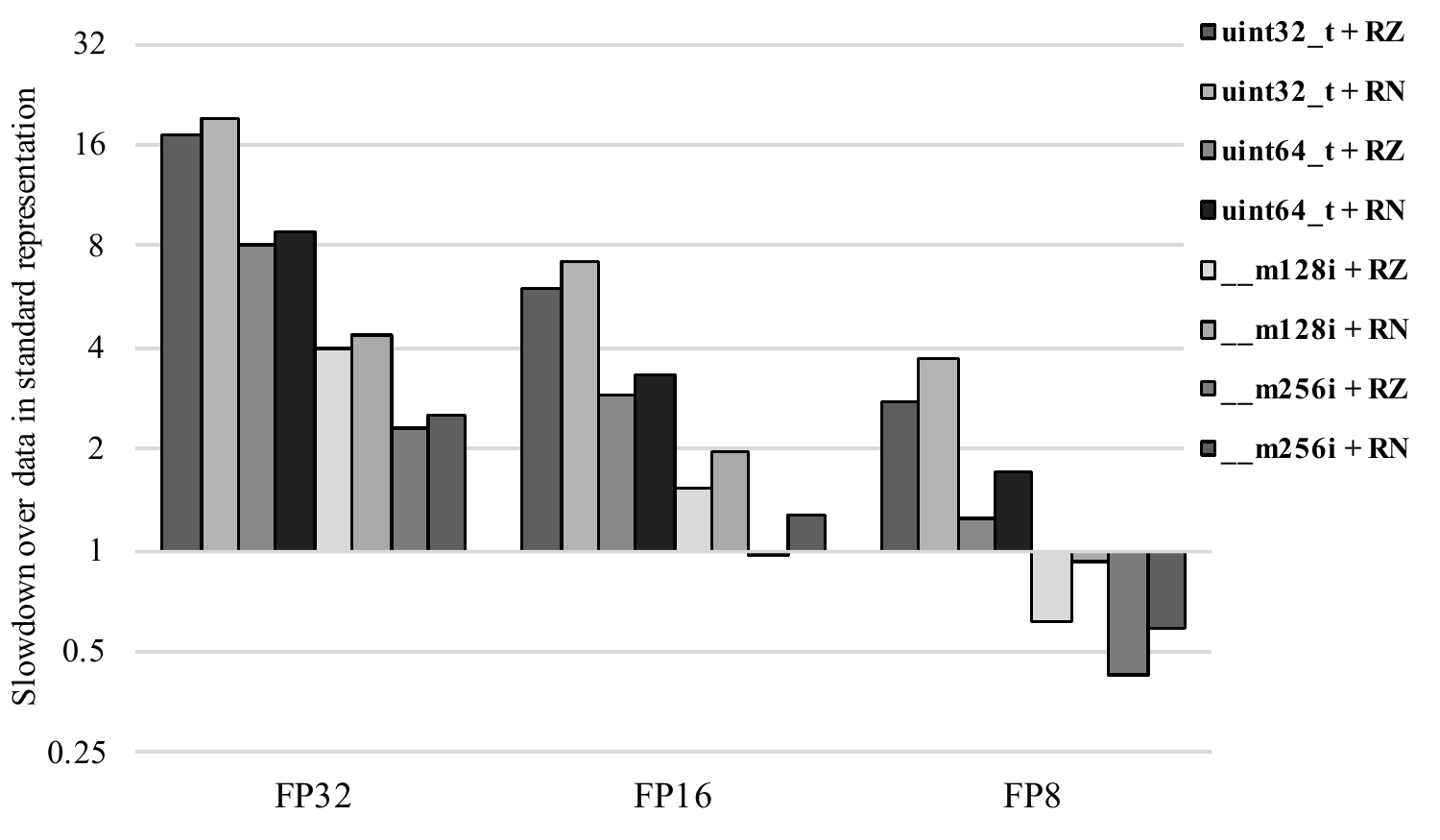}
  	\caption{Performance of BFP addition. RZ -- round towards zero; RN -- round to nearest.}
  	\label{add-perf}
  \end{figure}
  
  \begin{figure}[!bhpt]
  	\centering
  	\includegraphics[width=0.45\textwidth]{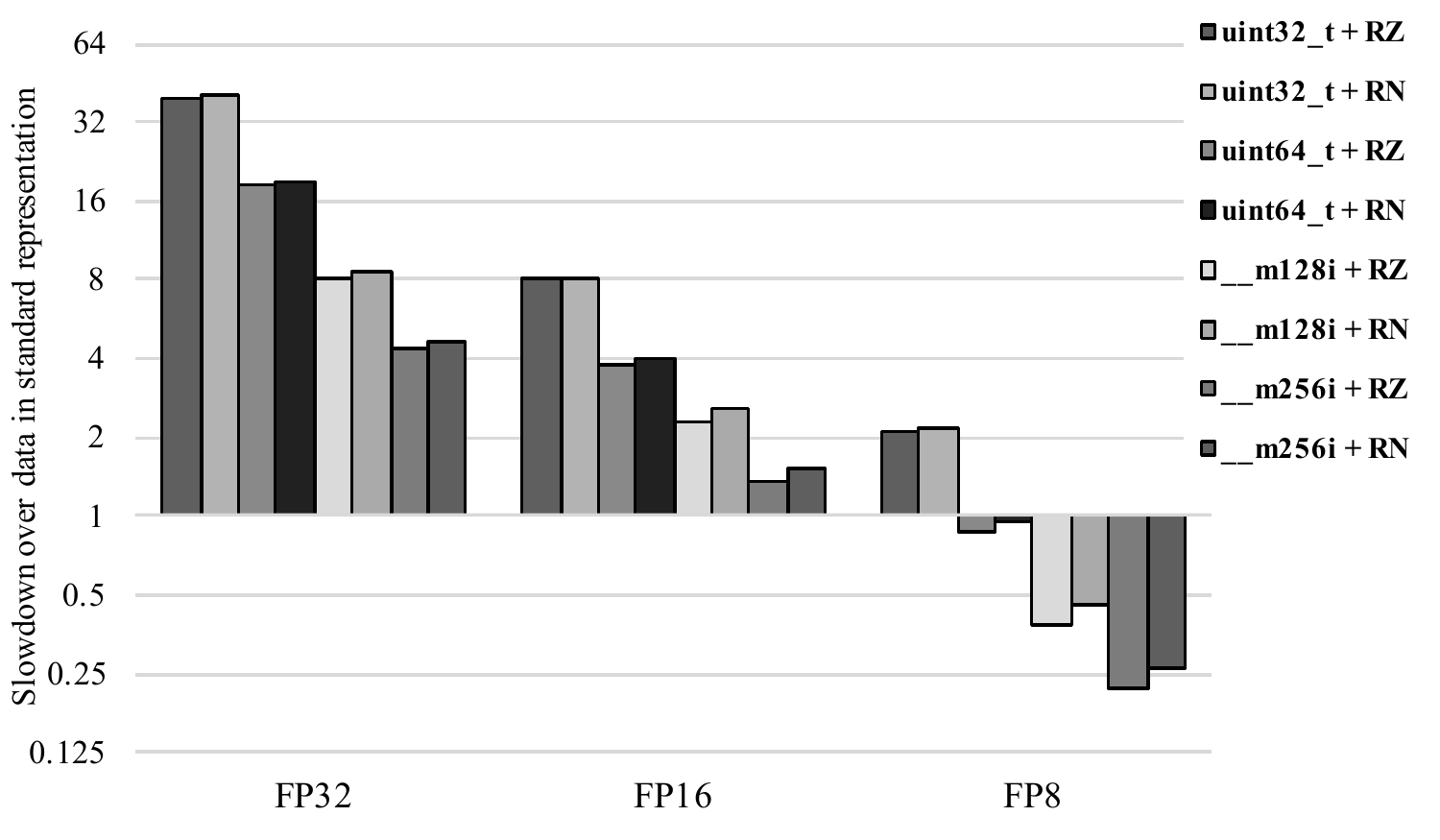}
  	\caption{Performance of BFP multiplication. RZ -- round towards zero; RN -- round to nearest.}
  	\label{mul-perf}
  \end{figure}

  \begin{figure}[!bhpt]
  	\centering
  	\includegraphics[width=0.45\textwidth]{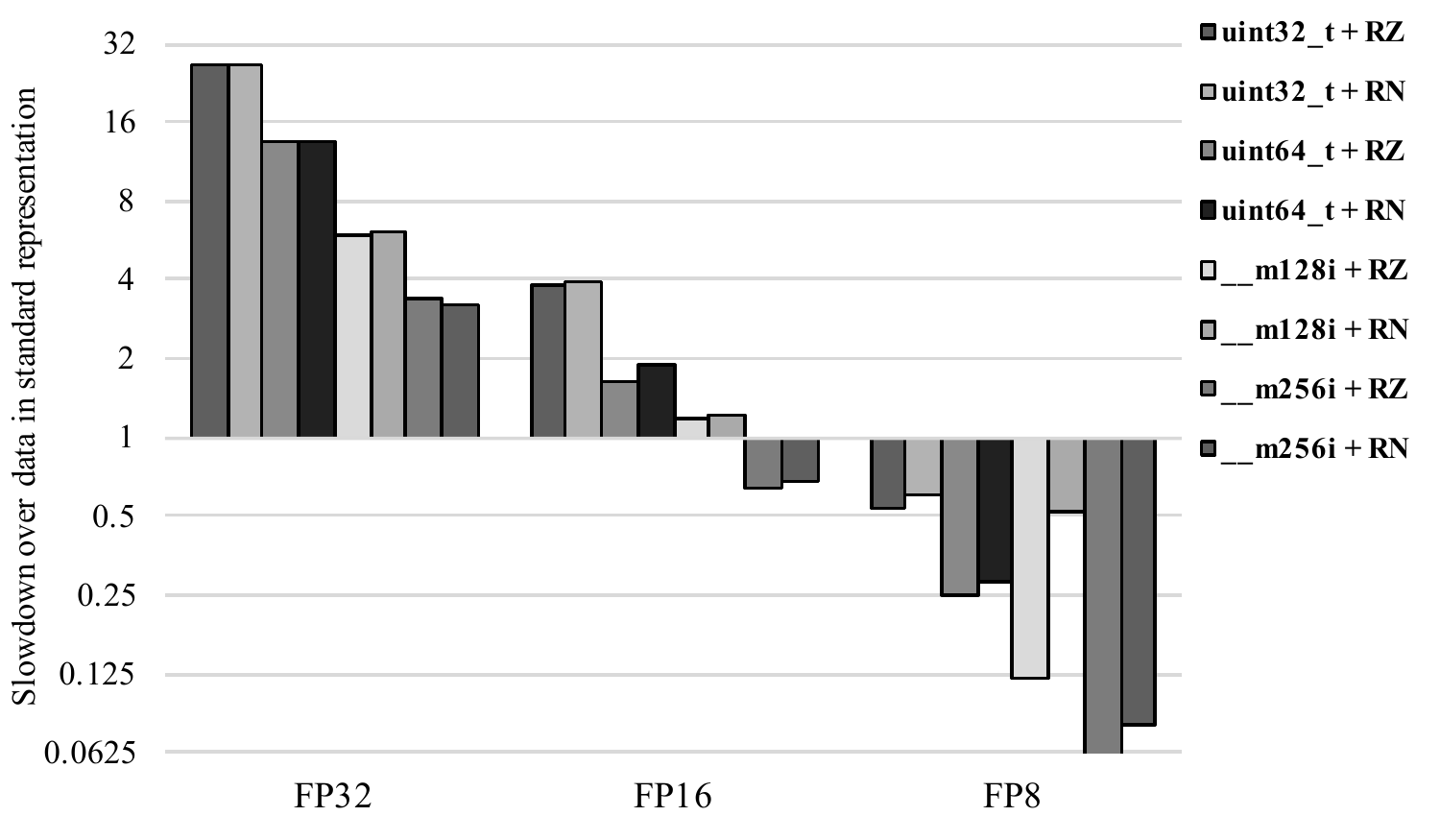}
  	\caption{Performance of BFP division. RZ -- round towards zero; RN -- round to nearest.}
  	\label{div-perf}
  \end{figure}
  
Figure \ref{add-perf} shows the performance of BFP addition. The performance of BFP multiplication and division is shown in Fig. \ref{mul-perf} and Fig. \ref{div-perf}, respectively. All the performance comparison is against the floating-point computations on data in the standard representation without using SIMD instructions. Without SIMD floating point units, our work shows another possibility to achieve SIMD floating point operations with integer computation units while with flexible precision. The performance results demonstrate that 1) with larger integer types, for example \_\_m256i, which allows more elements (256) being processed in parallel, the performance improvement is proportional to the width of the underlying integer types; 2) applying round towards zero (RZ) rather than round to nearest(RN) helps significantly improve the performance by reducing the width of significand in the intermediate results. In particular, for small floating point formats, our BFP multiplication and division can outperform the hardware counterpart greatly. As these formats are common in image processing and other fields where approximate computing is applicable, the great performance of our BFP vector operations with flexible precision makes our solution a feasible approach for approximate computing.

\section{Related Work}

SWAR~\cite{DBLP:conf/lcpc/FisherD98}, SIMD Within A Register, is the most closely related work to our bitslice floating point vector computation. SWAR uses logic operations to implement integer operations. Partitioned operations are the main focus of most of the hardware support for SWAR.
Instead of keeping the field size as the size of desirable data types, we use one single bit to hold one bit of a data element so that the number of elements processed in parallel is not decided by the width of data type but the max width of registers (including SIMD registers). Meanwhile, we use logic operations to implement the actual hardware logic for a single bit of the data element.

Lowering energy consumption is one of the major benefits of approximate computing. Disciplined approximate programming asks programmers to specify which parts of a program can be computed approximately. The approximate computation thus reduces the energy cost. An ISA extension is put forward to provide approximate operations and storage \cite{DBLP:conf/asplos/EsmaeilzadehSCB12}. With this extension, hardware has freedom to save energy at the cost of accuracy. Our customizable precision BFP vector types and related operations can serve as a software ISA for approximate computation.

Some programs may not need the dynamic range or the precision of FP arithmetic. For these programs, it is a general design practice to translate the floating-point arithmetic into a suitable finite fixed point presentation \cite{Tong:2000:RPO:340730.340741}. However, some programs may still require 6 bits or more in the exponent to preserve a reasonable degree of accuracy. In other words, these applications need more than the typical 32 bits of precision that fixed point arithmetic offers. Therefore, support for small, irregularly sized floating point makes our bitslice vector types a perfect fit for this kind of application.



\section{Conclusion}
We propose an entirely novel approach to vector computing based on
bitslice vector formats and building arithmetic operators from bitwise
instructions. This approach allows us to support a vector processing
model that can operate on data with an arbitrary number of bits. Thus,
we can create vectors of integer or floating point types of five,
nine, eleven or any number of bits. This ability to customize the
precision of vector data exactly to the application creates new
opportunities for optimization. In particular, it allows data
precision optimizations on general-purpose processors that were
previously available primarily on custom hardware. In addition,
matching precision to the application may reduce the memory footprint
of applications, which may in turn reduce memory traffic and the
energy required for data movement.

The complexity of the arithmetic operators is related to the number of
bits of precision in the data types. Our experiments show that for
larger precision, the costs of arithmetic operators becomes
prohibitive. However, for smaller data types the benefits of
exploiting bitwise parallelism across a vector of values can outweigh
the costs of bitwise arithmetic. To our knowledge we are the first to
propose and evaluate general-purpose bitslice vector representations.
We believe that it is a promising approach for approximate computing
using just enough precision.

\ifCLASSOPTIONcompsoc
  \section*{Acknowledgments}
\else
  \section*{Acknowledgment}
\fi

This work was supported by Science Foundation Ireland grant 12/IA/1381 and 10/CE/I1855 to Lero -- the Irish Software Research Centre (www.lero.ie).

\ifCLASSOPTIONcaptionsoff
  \newpage
\fi



%
\bibliographystyle{IEEEtran}
\bibliography{IEEEabrv,researchlib}

%
%
%
%
%




\end{document}